\documentclass[graybox]{svmult}
\usepackage{graphicx} 
\usepackage{amsmath, amsfonts, amsthm}
\usepackage{lmodern}
\usepackage[T1]{fontenc}
\usepackage{mathrsfs}
\usepackage{xcolor}
\usepackage{hyperref, caption}
\usepackage{subcaption}

\begin{document}

\title*{The Fleeting Laboratory: An Experimental Guide for Total Solar Eclipses}

\author{Suprit Singh and Bharti Arora}
\institute{Suprit Singh\orcidID{0000-0002-2530-3812} \at Indian Institute of Technology Delhi, Hauz Khas, New Delhi 110016 India \email{suprit@iitd.ac.in}\\
Bharti Arora\orcidID{0000-0002-1360-4853} \at Jaypee Institute of Information Technology, A 10, Noida, Uttar Pradesh 201309 India \email{bhartiarora612@gmail.com}
}
\maketitle

\abstract{Since times immemorial, total solar eclipses have inspired awe and wonder. In the modern scientific era, they have transformed into exclusive natural laboratories, offering fleeting but invaluable opportunities to study the Sun's faint outer atmosphere that is otherwise obscured by the intense glare of the photosphere. This unique vantage point has enabled revolutionary discoveries, from the identification of the element Helium and the first empirical validation of Einstein's General Relativity, to deciphering the corona's surprisingly high temperature. Today, ground-based eclipse experiments provide crucial data that complements and calibrates our space-based solar observatories, and offer high-resolution capabilities in the spatial, temporal as well as spectral domains. This chapter serves as a comprehensive guide detailing how to leverage modern observing equipments, detectors, and advanced computational techniques in image and data processing to conduct meaningful scientific investigations, bridging the gap between historical precedent and cutting-edge research.}

\section{The Evolution of Record Keeping}

\begin{figure}[h]
\begin{subfigure}{0.47\textwidth}
\centering
\includegraphics[height = 5cm]{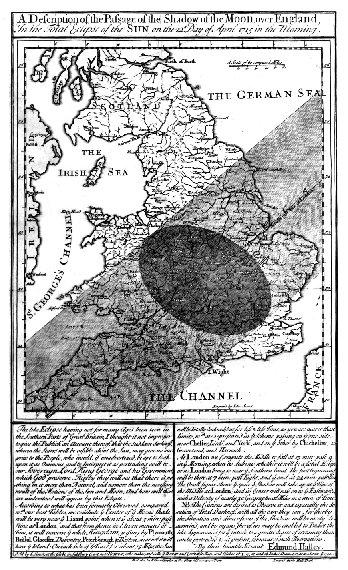} 
\caption{Edmond Halley's map of TSE 1715.}
\label{fig:subim1}
\end{subfigure}\hfill
\begin{subfigure}{0.47\textwidth}
\centering
\includegraphics[height = 5cm]{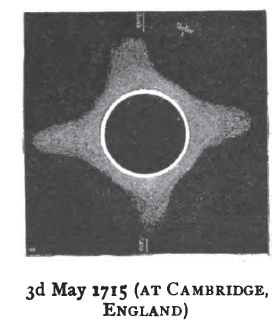}
\caption{A drawing of observations}
\label{fig:subim2}
\end{subfigure}
\par\bigskip
\begin{subfigure}{\textwidth}
\centering
\includegraphics[height = 5cm]{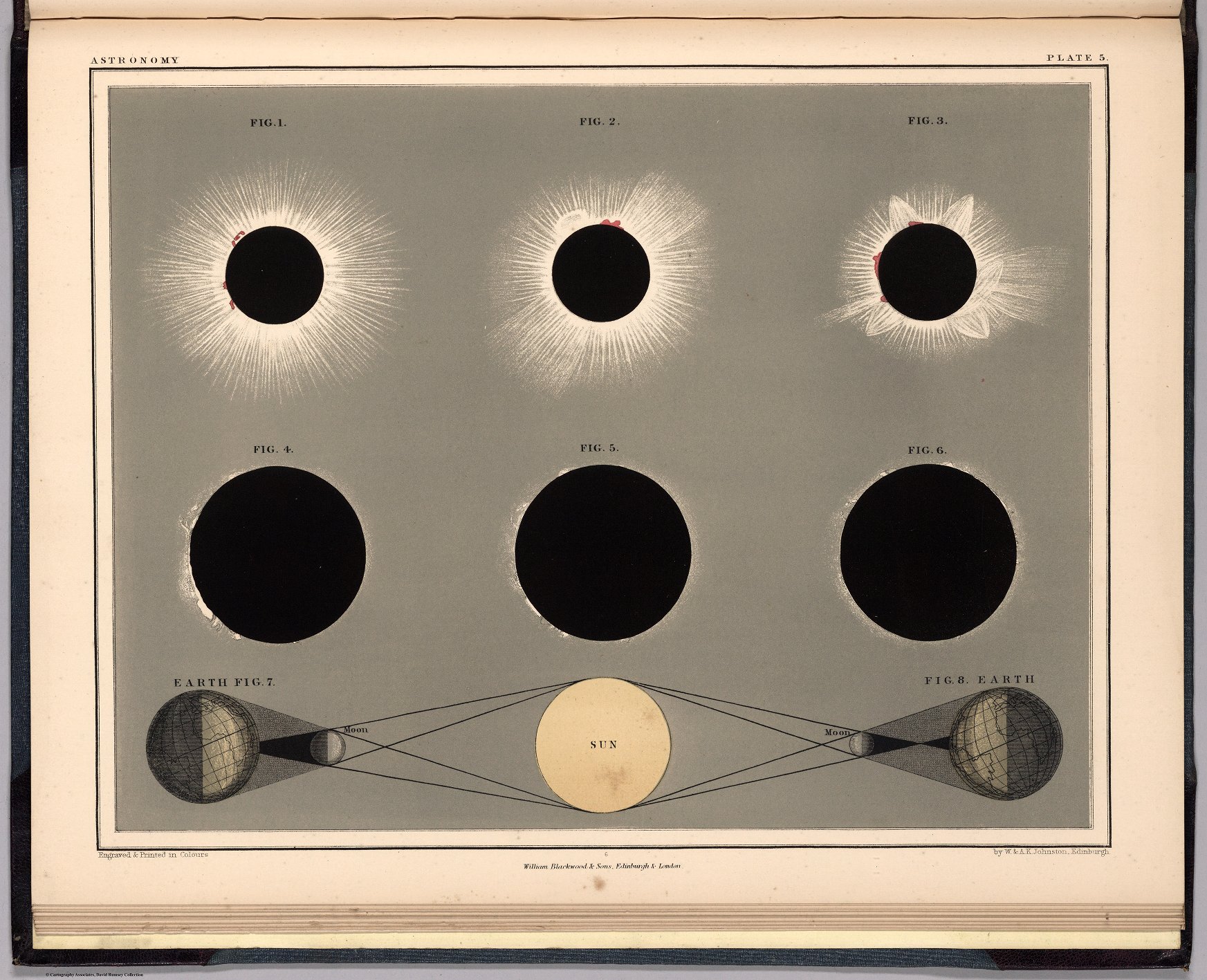}
\caption{An eclipse map from 1869 showing the solar corona and the prominences.}
\label{fig:subim3}
\end{subfigure}
\caption{Olden Records from the Total Solar Eclipses}
\label{fig:image1}
\end{figure}

We have amazing total solar eclipses (TSEs) on the Earth owing to the near-perfect match between the apparent sizes of our Moon and the Sun in the sky\cite{golub2010solar, 2001nsss.book.....G, held2005eclipses}. This allows the disc of the Moon on a New moon day to cover the glaring photosphere of the Sun, albeit, for a fleeing moment. The act reveals the faint and hidden outer atmosphere of the Sun in a sight unlike any other celestial phenomenon, and people have always tried to record these observations. The oldest record keeping dates as far as over 3000 BC in ancient Ireland, India, and China \cite{vaquero2009sun,guillermier1999total, 2006fmcs.book.....E}. The scientific age of eclipse related astronomy began with William Whiston and Edmund Halley's predictive maps of the TSE (Fig.~\ref{fig:subim1}\cite{Todd1894EclipseImage}) and the observations  (Fig.~\ref{fig:subim2}) \cite{Halley1715Map, 1996QJRAS..37..349C}) of the TSE of 1715 by astronomers from Cambridge, England. The drawings from the TSE 1869 show great details in terms of the prominences (Fig.~\ref{fig:subim3} \cite{johnston1869school}). Record keeping is of such an essential importance that even having a knowledge of an occurrence of an eclipse can help determine variations in the Earth's rotation over several thousand years\cite{stephenson1997heer}.

\begin{figure}[t!]
    \centering
    \includegraphics[scale = 0.05]{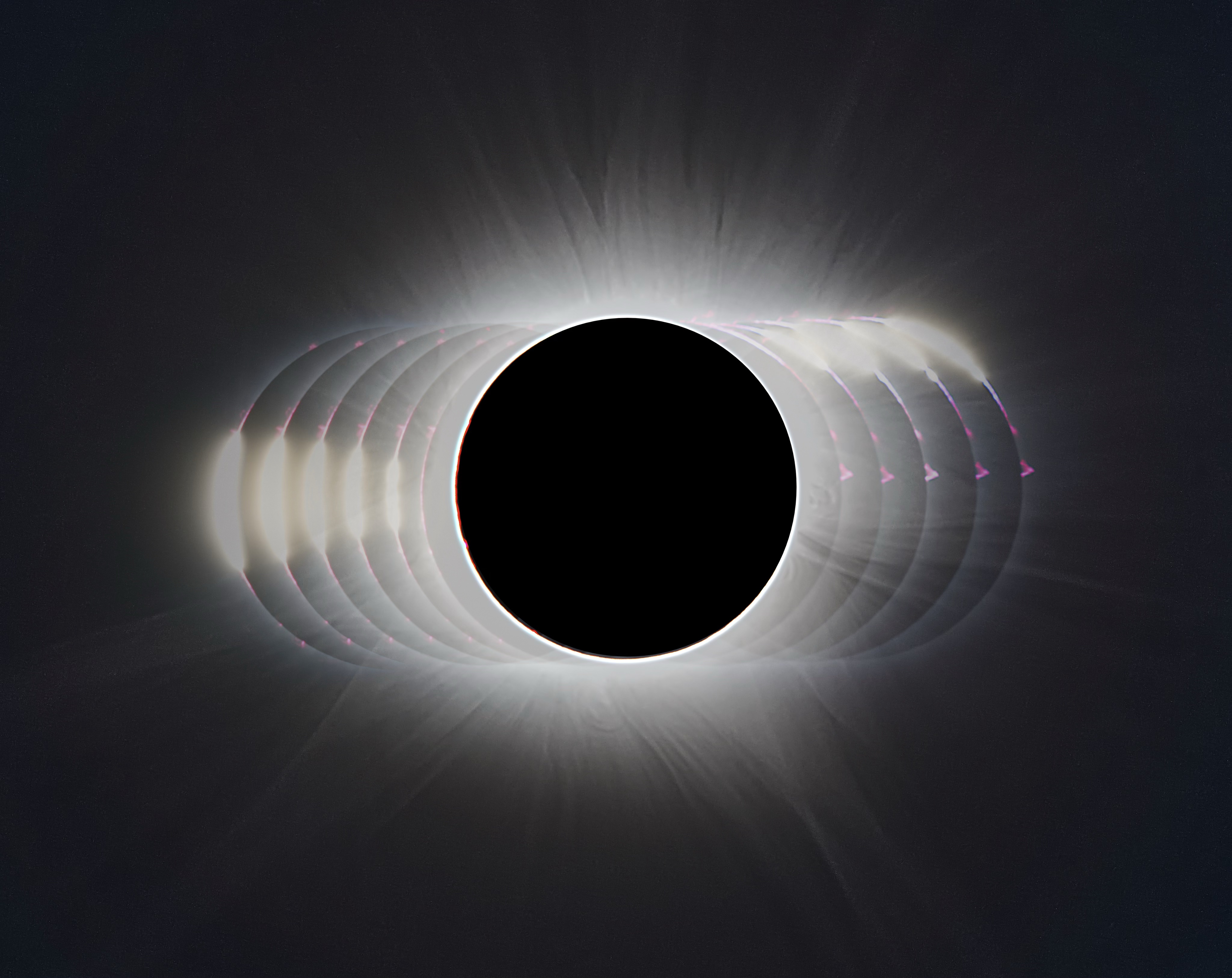}
    \caption{Moments during the Total Solar on April 08, 2024 captured at Dallas, Texas, US. As the disc of the Moon moves in to cover the Sun's photosphere, one sees the appearance of the Baily's beads, prominences and the solar corona. \copyright 2024 Suprit Singh and Bharti Arora.}
    \label{fig:image2}
\end{figure}

The key in taking these observations, processing of data, and gaining an understanding of the various physical phenomenon \cite{zirin1988astrophysics, foukal2004solar, aschwanden2009physics, Pasachoff:2009aa} has turned several epochs since then, to what is now possible with the current technology \cite{ozkan2007modern} (Fig. ~\ref{fig:image2}). This technological evolution, from Halley’s maps and hand-drawn prominences to high-resolution digital imaging, naturally brings forth a question. In an age where permanent solar observatories orbit our planet and fly through the solar wind itself, what role is left for the fleeting, Earth-bound eclipse experiments? As it turns out, the terrestrial vantage point remains not just relevant, but indispensable, in offering unique advantages in flexibility, innovation, and discovery.

\section{The Terrestrial Advantage}

Even with advanced capabilities of space-based telescopes, the ground-based research during solar eclipses remains critically important \cite{guillermier1999total}. While space telescopes excel in many areas of coronal research unhindered by atmospheric restrictions and conditions, their instruments are designed for specific purposes and their technology is locked in years before launch. The space-based coronagraphs also suffer from high scattering in the inner corona, thus they are not able to produce meaningful signal to noise ratio in the inner corona region ($\sim 1.5 \mathrm{R}_\odot$). However, the probability of scattered light from the Moon entering the telescope during the TSE is minuscule, given the Moon's (which acts as an occulter) vast distance from Earth. Ground-based coronagraphs installed on high mountains observing the low corona are limited by the sky brightness, and even occult the lowest corona which is the most critical region and their images are often noisy (thermal) with low signal to noise ratios.

Eclipse expeditions, in contrast, offer significant flexibility in allowing the scientists to use the latest equipment and frame observations around new theoretical ideas. This enables a wider variety of experiments, often using larger telescopes than are feasible to send into space. Importantly, eclipse research provides a much less expensive way to gather diverse chromospheric and coronal information than any space mission. These terrestrial events also serve as an invaluable testbed for new instruments and methods before they are committed to a costly space launch, cementing the role of eclipse observations as a vital and enduring tool in solar astronomy. There are several science goals that continue to be a part of the eclipse observations.

\begin{itemize}
    \item {\bf Merging Eclipse and Space Observations}: While instruments like the Large Angle and Spectrometric Coronagraph (LASCO) on NASA/ESA's SOHO spacecraft provide continuous views of the outer corona, their internal occulters block the crucial inner corona due to its high brightness. Ground-based eclipse observations perfectly bridge this gap, capturing the faint, dynamic structures of the corona closest to the solar limb that are otherwise obscured. This complementary data is vital for a complete understanding of coronal dynamics from the solar surface out into interplanetary space.

    \item {\bf Testing General Relativity at Eclipses:} A century after Arthur Eddington's pioneering observations during the 1919 total solar eclipse \cite{crelinsten2006einsteins, 1920RSPTA.220..291D} confirmed Einstein's theory of General Relativity, eclipses continue to offer unique opportunities for precise measurements of the deflection of starlight by the Sun's gravitational field. Modern experiments leverage advanced optics and detectors to refine these measurements, pushing the boundaries of our understanding of gravity\cite{2005hgse.book.....E, 1976AJ.....81..452T,Dittrich:2025aa, Goldoni_2020, 1970PhRvL..24.1377M}.
    
    \item {\bf Detecting Fine-Scale Coronal Dynamics:} The brief moments of totality provide an unparalleled opportunity to study the dynamic phenomena at diverse spatio-temporal scales down to milliseconds. This allows the detection and study fine-scale dynamic phenomena, such as propagating waves, small-scale jets, and plasma instabilities \cite{2002SoPh..207..241P, 2008ApJ...676L..73V, 2007ApJ...665..824P, 1994A&A...281..249K, Rusin:2008aa} which are crucial for understanding energy transport and heating mechanisms in the corona. 

    \item {\bf Predicting the Solar Corona:} Leading up to an eclipse, advanced computational models are used to predict the structure \cite{rusin2008comparing,Mikic:2018aa}and appearance of the solar corona based on magnetogram data from the Sun's surface. Eclipse observations provide the ultimate validation for these models, allowing comparison of the predictions with actual data and refine their understanding of the underlying physics governing the corona's shape and evolution.

    \item {\bf Heating the Corona:} The mystery of why the Sun's corona is millions of degrees hotter than its surface remains one of the most significant unsolved problems in solar physics \cite{2006SoPh..234...41K, 2002SoPh..207..241P, williams2004soho, 1999SoPh..188...89C}. Eclipse observations, especially through spectroscopic and imaging studies, provide crucial data on the temperature, density, and velocity of coronal plasma, helping to identify and characterize the energy dissipation mechanisms responsible for coronal heating. During totality, the corona's faint emission lines become observable. High-resolution spectroscopy at eclipses allows for detailed analysis of the chemical composition, temperature, density, and flow velocities of the coronal plasma. This includes studying the Doppler shifts of spectral lines to measure bulk flows and the broadening of lines to infer kinetic temperatures and turbulence.

    \item {\bf Electron Number Density \& Magnetic Field:} The scattering of photospheric light by free electrons in the corona produces polarized light, and measuring this polarization provides vital information about the physical conditions of the coronal plasma \cite{1996PASJ...48..545I, 2003ApJ...599..596R}. Polarimetric observations \cite{Hanaoka:2021aa, Hanaoka2024, Druckmuller2024, 2020ApJ...904..178B, 2023ApJ...946...14B}(measuring the degree and direction of mostly linear polarization across different wavelengths) during an eclipse are essential for determining the electron number density and inferring the strength and direction of the magnetic field in the corona.
    
    \item {\bf Topology of Magnetic Field Loops:} The magnificent structures of the solar corona, visible during an eclipse, are shaped by the Sun's complex magnetic field. Eclipse imaging, especially when combined with polarization measurements, helps map the topology of these magnetic field loops \cite{Boe2020, 10.1063/1.1618540}, providing insights into how magnetic energy is stored, released, and drives coronal phenomena like flares and CMEs. An important phenomenon that has eluded scientists for many decades is the acceleration of the solar wind. The magnetic topology of the Sun determines the speed of the solar wind whether slow and fast. Thus, understanding the magnetic topology of the Sun is crucial.
    
    \item {\bf Tracing Coronal Lines:} By carefully analyzing the intensity and morphology of various emission lines in the corona, we can trace specific plasma structures and differentiate between regions of different temperatures and densities. This helps in understanding the connectivity between different parts of the corona and its interaction with the underlying chromosphere and photosphere \cite{Boe2020}.
\end{itemize}

Achieving these goals requires an amalgamation of science and technology that goes in planning and taking observations during a total solar eclipse.

\section{Observing the Eclipse}

The intensity of a TSE event is so enthralling for observers that the build up to the time until totality is counted in days, then hours, then minutes and finally to seconds! The scientific observations need meticulous planning and execution. The best thing to do is to enjoy the whole thing visually for the first time.

\subsection{Placing yourself in the path of totality}

\begin{figure}
    \centering
    \includegraphics[width=0.65\linewidth]{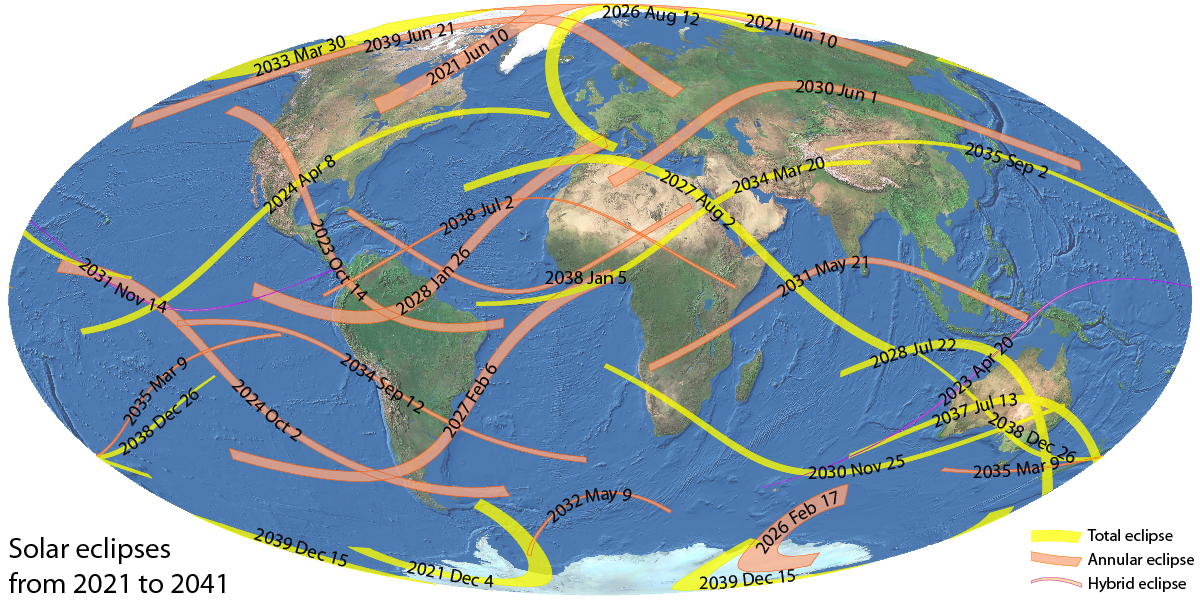}
    \caption{A world map depicting the tracks of solar eclipses for the period of 2021 to 2040 \cite{Zeiler_World_Eclipses_2040}.}
    \label{fig:map}
\end{figure}

While planning to observe an eclipse, the first and foremost decision is to review the path of totality (Fig.~\ref{fig:map}). It refers to the all the places through which the middle of shadow of the moon will pass through on the Earth during the eclipse. This is generally depicted to be the central line where the extent of spatial coverage of the lunar disc over the solar disc as well as the time of totality is maximum. Moving away from the center line reduces both the aspects. So the plan has to figure out an appropriate place on land in terms of accessibility, terrain, and the altitude of the Sun during totality. All of these factors weigh in equally as well as the need for favorable the weather conditions. These can only be judged somewhat from historical weather records of the location, and conditions on 2-3 days leading up to the eclipse. While accessibility to the places on the central line is subjective, the selection of a place based on terrain and altitude is important. For one, the view of the eclipse even for the partial phases has to be unobstructed \footnote{Altitude can be checked at \href{https://www.suncalc.org/}{https://www.suncalc.org/}.} if one has to record the photometric brightness of the solar disc for calibrations and subsequent coronal measurement, and secondly, the effects due to atmospheric seeing are minimal if the totality is high in the sky. Once the location is sorted out, the choice of equipment for the experiments depends on the scientific goals. We shall focus mainly on photometric studies, and then briefly comment on the polarimetric and spectroscopic observations.

\begin{figure}[t]
    \centering
    \includegraphics[width=0.6\linewidth]{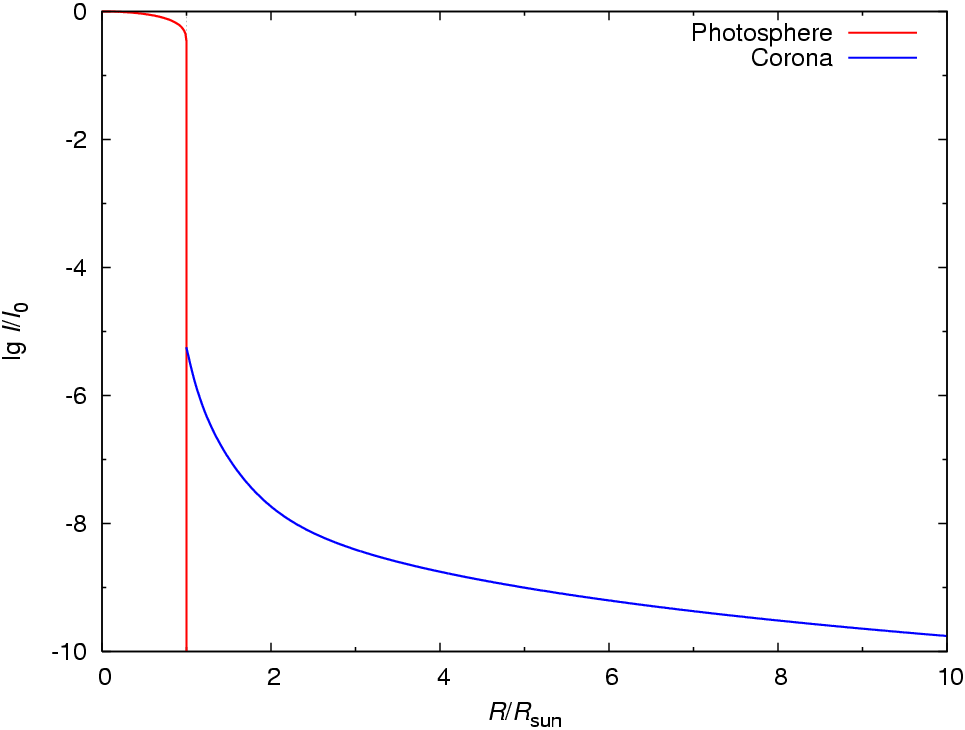}
    \caption[Brightness profile of the solar corona]{Brightness profile of the solar corona. Licensed under CC BY-SA 3.0. \protect\url{https://commons.wikimedia.org/wiki/File:Corona_Brightness_Profile.png}}
    \label{fig:corona_brightness}
\end{figure}

In order to collect the photometric data, the primary choice to be made in the equipment category is that of the camera and the telescope or the Optical Tube Assembly (OTA). The brightness of solar corona composed of three components \cite{golub2010solar, M.-Pasachoff:2017aa}: (i) K-Corona: photospheric light Thomson scattered by the free coronal electrons, (ii) F-Corona: light scattered by the interplanetary dust and (iii) E-Corona: emission from the coronal ions in highly ionized states (H$\alpha$, HeI$\mathrm{D}_3$, FeX, FeXIV, Ca XV etc.) in near UV-Visible-near IR spans a huge dynamic range being faint by five to nine orders of magnitude relative to the solar disc\cite{Liberatore:2022aa}. The K+F brightness goes down radially extending up to 6$\mathrm{R}_\odot$ and even more as show in Fig.~\ref{fig:corona_brightness}. This requires sensors with to have features such as an almost linear response, greater quantum efficiencies, large dynamic range, low dark noise, high resolution, large area to cover wide a field of view, and coupled with large and fast memory buffer that allows taking a significant number of raw exposures in short time giving an high temporal resolution. Having all these features and excelling in each of them sounds like an ideal imaging system.

\subsection{The Camera}

This is where the modern CMOS sensors have turned the tables which is also why we living in exciting times for eclipse based experiments. An extent of these features is found in the consumer grade and easily accessible mirrorless cameras. These are like digital single-lens reflex (DSLR) cameras sans the mirror, hence the optical view finder is replaced by an electronic view finder and the back of the camera displays provide an almost zero latency live view. The cameras are therefore light weight and live view allows easy focus adjustments during the captures. These cameras, unlike dedicated astronomy, cameras are also equipped with large RAW buffer, dual card slots for data redundancy, and can be triggered with intervalometer or scripts when tethered with a computer. 

There are some points worth mentioning which can be drawbacks in some situations. These cameras are one shot color cameras since they are made for usual daylight photography. The sensors have a Bayer filter array on top, mostly in the RGGB configuration, to be able to record color information simultaneously in the three RGB channels. This makes the task of photometry a bit non-trivial since the individual sensitivities of these filters and their bandwidths have to be taken into account which can require characterizing a camera for calibrations. It also does not yield the true pixel resolution since the Bayer array leaves gaps in the individual RGB channels which are filled by interpolation also referred to as debayering. Further, an experiment may require full bolometric measurement of flux over the optical bandwidth. In this case, the only option is to create a synthetic luminance channel by combining the three RGB channels with appropriate weights. But when the data is recorded through the RGB channels, it does have some spectral information in a way, and brightness in these individual channels can be useful for some science goals, for example, in the separation of K and F corona. Finally, these mirrorless cameras are not cooled and hence cannot capture the images under a controlled environment of a fixed sensor temperature, so dark subtraction or calibration is not always spot on as dark noise varies with temperature. However, the dark current is small and is not found to vary much with exposure times in the the modern sensors, and thus this limitation is not much of an issue. Unfortunately, there are also some manufacturer level issues in the cameras from all the consumer brands which can cause incurable artifacts and many of these have been documented \cite{ShelleyCameraSummary}. As of writing this article, the Sony A7RV (61MP sensor with 3.76$\mu$m pixel pitch) has been listed to be the only mirrorless camera that is free from most of the issues. For spectroscopy, it is generally preferable to use a monochrome sensor which is available only in dedicated astronomy cameras. A way to get faster speeds from the dedicated astronomy cameras is to use them tethered to a computer so that the fits files can be transferred directly and fast, and use a region of interest (ROI) mode during captures. It sacrifices some field of view for speed.  

\subsection{The Optical Tube Assembly (OTA)}

A camera has to be coupled to the right OTA for gathering the data. The selection of the OTA therefore goes hand in hand with that of the camera. Since we are limited in the choice of the camera, the option of an OTA is constrained accordingly. There are several aspects that are to be taken care of here: (i) resolution in terms of the pixel scale \cite{AstronomyToolsFOV}, (ii) atmospheric seeing, and (iii) portability since the eclipses often happen at places where travel is necessary. With the advent of modern high resolution ($\sim 61$MP, 9504 x 6336 pixels, 3.76$\mu$m) sensors, the resolution scales of $\sim 1.3''$ per pixel have become possible to achieve with scopes of 80mm aperture and focal ratios of F7.5. This yields a focal length of 600mm for a Full Frame (sensor size: 35.7 $\times$ 23.8 mm) camera and a FOV of $3.42^\circ \times 2.28^\circ$ enough to cover out to $\mathrm{R}_\odot$. One more constraint that is placed on the OTA is the flat field coverage up to an image circle of $\phi$44 mm to be able to utilize the Full Frame (FF) cameras. Below this, one has to use an APSC sized sensor ($\approx1.5$ times crop of the FF) or even smaller, and hence the FOV will be cropped accordingly. Once these numbers are in place, the type of OTA can be a refractor or a reflector based on the requirement of portability. Refractors are usually preferred in this regard, and also tend to be more resilient to temperature changes although focus has to be checked and set regularly during the eclipse. If a wide-band (over full optical spectrum) the data is being recorded, an apochromatic refractor (Flourite doublets, or better a triplet) should be used to keep the chromatic aberrations in check. Newtonian and other reflectors do not suffer from this issue, but need coma correctors for a flat field coverage. Also, the reflectors work better in the near-IR experiments. Ref. \cite{Molnar:2025aa} used a Skywatcher Newtonian for spectroscopic observations of the He I 1083 nm wavelength during the TSE2024.

\subsection{Tracking the Sun}
We need to track the Sun and the Moon for the whole duration of the eclipse. This is often automated by the use of Go-To mounts with motorized movement and in-built sky models for pointing. These mounts can also be connected and controlled from a PC along with a dedicated hand-controller. There are two types of mounting configurations in use, the Altitude-Azimuth(Alt-Az) and the equatorial configurations. The first works on the basis of Horizontal coordinate system and requires simultaneous movement in both directions (both motors in the two axes running) to track the object. The equatorial configuration uses celestial coordinates, Right Ascension-Declination (RA-Dec) system, in which the mount has to be polar aligned, so that once the object has been located, the Dec remains fixed and that only movement that happens, occurs around the polar axis. There is also no field rotation in this case. For the Alt-Az configuration, a de-rotator is needed in the imaging train to counter the field rotation. Modern Harmonic drive mounts offer a high weight to payload capacity. These are very portable and best for eclipse expeditions. Polar alignment for working in the equatorial configuration can be done either night before, early morning before twilight, or roughly with a compass that allows dialing in the magnetic declination correction based on the location. 

Tracking is usually good if the polar alignment is excellent. To make it best, a solar guider can be used for auto-guiding. There is an excellent one from Hinode \cite{HutechHinode} which is a standalone auto-guider and does not require a PC to work and works even with a rough polar alignment. The auto-guider comes with a controller, guiding assembly (optics, filter, and sensor are integrated in a single package), solar finder for lining up the Sun, and connects to the mount via the ST4 cable giving corrections upon drifting. 

\subsection{Achieving Focus}

There is a noticeable change in the environment as well temperature drop as totality approaches. The cooling can also clear up the clouds in the sky leading to good observing conditions. It is thus advisable to use motorized focusers with an external temperature sensor taped to the underneath of the telescope tube towards the front (closest to the lens assembly). The focus change versus temperature can been calibrated for the OTA. During the eclipse, the direction of focus change is the opposite to what you would expect if adjusting for expansion/contraction of the tube. The change in refractive index of the glass has more impact on focus position and it is in the opposite direction to tube contraction \cite{Phil2024}. It is also important to note that if you are changing filters or anything in the imaging train that changes the path difference one should refocus after the change before taking more exposures. 

\subsection{Image Acquisition}

An eclipse is a fast-paced event with several phases taking place in quick succession, the acquisition of data during a TSE is unlike any other situation \cite{2007tseh.book.....M}. The partial phases are most easiest to record and are done periodically, with a safe solar filter in place at the objective (front) end of the OTA. There are several solar films available. Baader OD 5.0 has an even white light response over the optical bandwidth. For less attenuation, there is also a Baader OD3.8 available but it is only for photographic purpose and \emph{not for visual use}, while BaaderOD5.0 can be used for visual purposes as well. This is important point to note for safety, and will add here that while you are doing eclipse observations, it is \emph{essential} to have some solar glasses with you. Also the Baader OD5.0 to be used for imaging should be well inspected for perforations before use. A total solar eclipse has 4 contact points. These are the moments when the Moon begins to cover the Sun (first contact), the moment the Sun is fully covered (second contact), the moment totality ends and the Sun reappears (third contact), and the moment the Moon completely moves away from the Sun (fourth contact). Totality, the brief period of total darkness, occurs between the second and third contacts. 

\begin{enumerate}

    \item [C1] First Contact: The partial eclipse begins as the Moon's edge makes its first tangential touch with the Sun's edge. It looks like a small bite has been taken out of the Sun.

    \item [C2] Second Contact: The moment the Sun is completely covered by the Moon, and totality begins. This is heralded by the Baily's Beads \cite{1836MNRAS...4...15B} and diamond ring effects.

    \item [C3] Third Contact: The instant the Moon starts to move away from the Sun, and totality ends. Again, the diamond ring effect may be visible as the first light of the Sun appears.

    \item [C4] Fourth Contact: The final moment when the trailing edge of the Moon moves completely off the Sun, and the eclipse concludes. 

\end{enumerate}

The adrenaline-pumping action begins just before C2 and a bit after C3 and several events take place in quick succession in fleeting moments. Just before C2 (T-0 seconds) when the disc of the moon is about to cover the photosphere of the Sun, typically, T-60s: before you start to see the Shadow bands, T-30s: put your hands on the camera filter, T-20s: remove camera filters, you see diamond ring, T-15s: the Baily's beads, T-2s: the chromosphere and the prominences become visible and T-0s: the glasses are off, T+15s: Max eclipse, and you have moments until C3 the third contact when it is all over. C3+15s: Replace Camera filters. C3+25s: observe shadow bands. C3+45s: the umbra exits. C3+60s: the partial phase begins in reverse. The next checkpoint will be at C4 when it all ends and you begin looking forward to the next expedition. The very important thing is practice, practice and practice before the actual event. Even practicing it raises the adrenaline for sure. These numbers are just for giving you the idea, and do vary for every eclipse as well as the location, so use an accurate eclipse timer with proper location set using the GPS for getting the actual times. 

Since the whole event has a high dynamic range of varying brightness, naturally a varying set of exposures is necessary. The same is also necessary while taking the data of the Solar corona during totality, as it itself has a high dynamic range in terms of the radially varying brightness going from bright inner corona near the solar limb out to the fainter outer corona for several solar radii. For any High Dynamic Range (HDR) composite image, we take bracketed exposures and then compose them into a single image. The bracketed exposure timings (say, number \#M of them) can range from 1/800th of a second to 1-2s (to get the earthshine) depending upon the gain of the sensor (kept at ISO 100 usually, lowest for maximum dynamic range of the sensor), and the aperture in this case is fixed by the OTA. In order to increase the signal to noise ratio (SNR) we take a several images or light frames (say, number \#N) for each exposure time for the given set of \#M bracketed exposures mentioned above and then stack (or co-add) them. This can be achieved by repeating the bracketed exposures \#N times yielding (\#N $\times$\#M) number of total images. The pre-processing routine here is to first stack the frames with the same exposure timings thus yielding \#M stacked images each pertaining to different exposure times. These \#M images are then combined to obtain a final HDR image using appropriate algorithms. There are reports that with the modern sensors and their huge dynamic range available at the lowest gain values, one need not take bracketed exposures. We can simply take a large number of frames at the same exposure keeping the back of the camera histogram around the middle such the the data is neither underexposed or overexposed. Stacking these will increase the SNR significantly and with high DR of the sensor, the data can be pulled out by appropriate stretching. 

There are eclipse calculators \cite{JubierCircumstancesCalc} available that compute the timings of the events during a TSE and also exposure times \cite{JubierExposureCalc}. There are also programs that can automate the data acquisition run \cite{JubierMaestro}. But these have their own limitations in terms of the computer architecture they can be used on, and suitability for the kind of scientific experiment to be performed. There are also timers (LRTimelapse Pro Timer 3) that can send signals to the cameras for obtaining exposures as needed. These are crucial for data acquisition during the TSEs as taking exposers manually can be erroneous during the already pumped up event.

Finally, the light or science frames (data) have to be calibrated before stacking and for this, appropriate dark and flat frames are also taken. The darks can be taken before C2 and after C3 in the gaps between partial eclipse exposures. The exposure time of the dark frames has to match the exposure time of the light frames from which they are to be subtracted. Once again the same procedure as for the light frames can be followed, that is, repeat the (\#M) bracketed exposures several times but with the telescope lid closed. Stack dark frames of each exposure time to get \#M Master dark frames. It is assumed that the variation of the dark current is not significant with the temperature changes, and a Master dark from these works well for calibration. Similarly, Flats can be taken after C3 so as to correct for the uneven illumination in the field and across the field, and a Master flat (stacked flat frames) is then used for calibration. It is important to use the flat panels that are evenly lit and also have a flat spectral brightness for obtaining the flats. The exposure time of the flat frames is chosen such that the back of the histogram of the camera is in the middle, giving neither under nor over exposed images. If needed, the corresponding dark frames for these flat frames can also be taken to correct the flats for the dark current/noise.

\subsection{Other Equipments}

Depending upon the kind of experiment being performed, there can be other sets of equipments to be added and controlled in the imaging train. As an example, for obtaining the polarization data \cite{Hanaoka:2021aa, Hanaoka2024, Druckmuller2024, Edwards:2023aa}, one can use a set of polarizers installed in the filter wheel between the camera and the OTA, use a rotating polarizer assembly, or use special cameras which have an array of polarizers much like the RGGB bayer array, over the pixels of an otherwise monochrome sensors which can simultaneously record polarization at different angles. We used a symmetric triplet of polarisers at 0, 60 and -60 degrees to the horizontal axis of the camera sensor installed in a filter wheel to obtain the polarimetric data with a Nikon Z7 camera in RGB during the TSE2024. This gives an additional advantage of having a spectral information along with the polarization, albeit in the wider bands and not the spectral lines which a dedicated spectro-polarimeter would yield. Nevertheless, having some information is better than having no information at all.

In order to study the emission E-corona, one has to do spectroscopy \cite{2003ApJ...599..596R}. In this case one can use a slit-less spectrograph \cite{Voulgaris:2012aa, Voulgaris:2022aa}, Fabry-Perot filter\cite{2008ApOpt..47.5744N}, a grating or a grism at the objective of the lens or a refractor, or work with a spectrograph that has a slit and then scan the corona with the same \cite{Muro:2023aa}. Within a spectrograph, wavelengths can be selected either by rotating the grating, or by having more than one detectors at different angles thus imaging multiple channels simultaneously. In principle, the E-corona can also be observed to a large extent by using a narrow band $(\sim$ 1\r{A}) filter. Such filter images can capture the different regions of the corona simultaneously. The choice of setup is dictated by the goals of the experiment and the spatial, temporal, and spectral resolution required therein.

\subsection{Essential Accessories}

Since the eclipse expeditions and experiments are unlike any other, apart from the experimental equipment, there are several other useful items to be kept on the side. We list them here for completeness. These are a notebook and digital voice recorder for note taking, powering equipment (batteries, extension cords), powered USB Hubs, extra memory cards, velcro strips, duct and masking tapes for emergency repairs, eclipse glasses, headlamp, tools (screw drivers, Allen keys, a small wrench, tweezers)  for minor repairs, camera adapters and an extra set of cables, large rain covers, compass with magnetic declination correction, and a GPS device.

\section{Wait for the next}

An eclipse is not even over when talks about the next one begin already. The countdowns are in place, and a lot of science is always waiting to be explored. New questions arise, the technology keeps advancing, ideas keep getting ignited. In terms of instrumentation, until now with the last eclipse being TSE2024, several arenas have been explored from photometry, spectroscopy, as well as polarization. The next leap will be in doing spectro-polarimetry during the eclipses. The setup has to be made portable, calibrated, adaptable to smaller telescopes. The polarization of emission lines is key to understanding the coronal magnetic fields and the turbulent outer atmosphere of the Sun. 

Total Solar Eclipses are unique blend of savoring the beauty of nature and understanding how it works.


\bibliographystyle{spphys}
\bibliography{ISSAC_SS}

\begin{thebibliography}{10}
\providecommand{\url}[1]{{#1}}
\providecommand{\urlprefix}{URL }
\expandafter\ifx\csname urlstyle\endcsname\relax
  \providecommand{\doi}[1]{DOI \discretionary{}{}{}#1}\else
  \providecommand{\doi}{DOI \discretionary{}{}{}\begingroup \urlstyle{rm}\Url}\fi

\bibitem{golub2010solar}
L.~Golub, J.M. Pasachoff, \emph{The Solar Corona}, 2nd edn. (Cambridge University Press, 2010)

\bibitem{2001nsss.book.....G}
L.~{Golub}, J.M. {Pasachoff}, \emph{{Nearest star : the surprising science of our sun}} (2001)

\bibitem{held2005eclipses}
W.~Held, \emph{Eclipses: 2005--2017: A Handbook of Solar and Lunar Eclipses and Other Rare Astronomical Events} (Floris Books, Edinburgh, 2005).
\newblock Translated from the German work \textit{Astronomische Sternstunden}, published by Verlag Freies Geistesleben, 2005.

\bibitem{vaquero2009sun}
J.M. Vaquero, M.~V{\'a}zquez, \emph{The Sun Recorded Through History: Scientific Data Extracted from Historical Documents}, \emph{Astrophysics and Space Science Library}, vol. 361 (Springer, New York, 2009)

\bibitem{guillermier1999total}
P.~Guillermier, S.~Koutchmy, \emph{Total Eclipses: Science, Observations, Myths and Legends} (Springer/Praxis, London, 1999).
\newblock Translated by Pierre Guillermier from the French work \textit{Eclipses Totales: Histoire, D{\'e}couvertes, Observations} (Masson, 1998)

\bibitem{2006fmcs.book.....E}
F.~{Espenak}, J.~{Meeus}, \emph{{Five millennium canon of solar eclipses : -1999 to +3000 (2000 BCE to 3000 CE)}} (2006)

\bibitem{Todd1894EclipseImage}
M.L. Todd.
\newblock {Solar eclipse 1715May03-Cambridge England}.
\newblock Image from \textit{Total Eclipses of the Sun} (1894). Retrieved from Wikimedia Commons (1894).
\newblock URL: \url{https://commons.wikimedia.org/wiki/File:Solar_eclipse_1715May03-Cambridge_England.png} [Public Domain]

\bibitem{Halley1715Map}
E.~Halley.
\newblock A description of the passage of the shadow of the moon, over england.
\newblock Broadside map (1715).
\newblock Map published for the total solar eclipse of 1715 May 3 (Julian calendar 22 April).

\bibitem{1996QJRAS..37..349C}
A.~{Cook}, \qjras \textbf{37}, 349 (1996)

\bibitem{johnston1869school}
A.K. Johnston, \emph{School Atlas of Astronomy: Comprising, in Twenty-one Plates, a Complete Series of Illustrations of the Heavenly Bodies} (William Blackwood and Sons, Edinburgh, 1869)

\bibitem{stephenson1997heer}
F.R. Stephenson, \emph{Historical Eclipses and Earth's Rotation} (Cambridge University Press, Cambridge, 1997)

\bibitem{zirin1988astrophysics}
H.~Zirin, \emph{Astrophysics of the Sun} (Cambridge University Press, Cambridge, 1988)

\bibitem{foukal2004solar}
P.~Foukal, \emph{Solar Astrophysics}, 2nd edn. (Wiley-VCH, Weinheim, 2004)

\bibitem{aschwanden2009physics}
M.J. Aschwanden, \emph{Physics of the Solar Corona: An Introduction with Problems and Solutions} (Praxis Publishing, Chichester, UK, 2009).
\newblock 3rd printing

\bibitem{Pasachoff:2009aa}
J.M. Pasachoff, Nature \textbf{459}(7248), 789 (2009).
\newblock \doi{10.1038/nature07987}.
\newblock \urlprefix\url{https://doi.org/10.1038/nature07987}

\bibitem{ozkan2007modern}
M.T. Ozkan, et~al., in \emph{Modern Solar Facilities--Advanced Solar Science}, ed. by F.~Kneer, K.G. Puschmann, A.D. Wittmann (Universit{\"a}tsverlag G{\"o}ttingen, G{\"o}ttingen, 2007), pp. 201--204

\bibitem{crelinsten2006einsteins}
J.~Crelinsten, \emph{Einstein's Jury: The Race to Test Relativity} (Princeton University Press, Princeton, 2006)

\bibitem{1920RSPTA.220..291D}
F.W. {Dyson}, A.S. {Eddington}, C.~{Davidson}, Philosophical Transactions of the Royal Society of London Series A \textbf{220}, 291 (1920).
\newblock \doi{10.1098/rsta.1920.0009}

\bibitem{2005hgse.book.....E}
D.S. {Evans}, K.I. {Winget}, \emph{{Harlan's Globetrotters: The Story of an Eclipse}} (2005)

\bibitem{1976AJ.....81..452T}
{Texas Mauritanian Eclipse Team}, \aj \textbf{81}, 452 (1976).
\newblock \doi{10.1086/111906}

\bibitem{Dittrich:2025aa}
W.A. {Dittrich}, D.G. {Bruns}, R.~{Berry}, K.~{Carrell}, D.~{Smith}, A.D.P. {Smith}, D.~{Borrero-Echeverry}, G.~{Kinne}, J.M. {Izen}, H.~{Hill}, G.~{Mulder}, J.J. {Rembold}, C.~{Delgado}, A.E. {Hornbeck}, S.A. {Jeffe}, J.R. {McSorley}, O.E. {Schutz}, M.~{Strate}, E.~{Matin}, J.~{Kinder}, P.~{Poncy}, C.~{Freels}, J.~{Benitez-Flores}, R.~{Smith}, B.~{Bauer}, C.~{Rajendram}, S.~{Leathers}, L.~{Fenstemacher}, M.P. {Clark}, E.~{Kempe}, T.~{Slaght}, K.~{Webb}, C.~{Bradley}, S.~{Plascencia}, G.~{Le}, A.C. {Moon}, Y.~{Choi}, A.~{Tom}, S.~{Youngquist}, K.~{Castaneda}, N.~{Marichalar}, I.~{Muench}, C.~{Nash}, R.~{Brown}, J.~{Obermiller}, G.~{Vetters}, E.~{Singh}, in \emph{Bulletin of the American Astronomical Society}, vol.~56 (2025), vol.~56, p. 2024n9i040.
\newblock \doi{10.3847/25c2cfeb.7f092727}

\bibitem{Goldoni_2020}
E.~Goldoni, L.~Stefanini, Physics Education \textbf{55}(4), 045009 (2020).
\newblock \doi{10.1088/1361-6552/ab8778}.
\newblock \urlprefix\url{https://doi.org/10.1088%2F1361-6552%2Fab8778}

\bibitem{1970PhRvL..24.1377M}
D.O. {Muhleman}, R.D. {Ekers}, E.B. {Fomalont}, \prl \textbf{24}(24), 1377 (1970).
\newblock \doi{10.1103/PhysRevLett.24.1377}

\bibitem{2002SoPh..207..241P}
J.M. {Pasachoff}, B.A. {Babcock}, K.D. {Russell}, D.B. {Seaton}, \solphys \textbf{207}(2), 241 (2002).
\newblock \doi{10.1023/A:1016297800478}

\bibitem{2008ApJ...676L..73V}
T.~{Van Doorsselaere}, V.M. {Nakariakov}, E.~{Verwichte}, \apjl \textbf{676}(1), L73 (2008).
\newblock \doi{10.1086/587029}

\bibitem{2007ApJ...665..824P}
J.M. {Pasachoff}, V.~{Ru{\v{s}}in}, M.~{Druckm{\"u}ller}, M.~{Saniga}, \apj \textbf{665}(1), 824 (2007).
\newblock \doi{10.1086/519680}

\bibitem{1994A&A...281..249K}
S.~{Koutchmy}, M.~{Belmahdi}, R.L. {Coulter}, P.~{Demoulin}, V.~{Gaizauskas}, R.M. {MacQueen}, G.~{Monnet}, J.~{Mouette}, J.C. {Noens}, L.J. {November}, \aap \textbf{281}(1), 249 (1994)

\bibitem{Rusin:2008aa}
V.~Ru{\v s}in, M.~Druckm{\"u}ller, M.~Minarovjech, M.~Saniga, Astrophysics and Space Science \textbf{313}(4), 345 (2008).
\newblock \doi{10.1007/s10509-007-9686-2}.
\newblock \urlprefix\url{https://doi.org/10.1007/s10509-007-9686-2}

\bibitem{rusin2008comparing}
V.~Ru{\v{s}}in, et~al., in \emph{Transactions of the American Geophysical Union} (2008), Fall Meeting.
\newblock Abstract SH13B-1524

\bibitem{Mikic:2018aa}
Z.~Miki{\'c}, C.~Downs, J.A. Linker, R.M. Caplan, D.H. Mackay, L.A. Upton, P.~Riley, R.~Lionello, T.~T{\"o}r{\"o}k, V.S. Titov, J.~Wijaya, M.~Druckm{\"u}ller, J.M. Pasachoff, W.~Carlos, Nature Astronomy \textbf{2}(11), 913 (2018).
\newblock \doi{10.1038/s41550-018-0562-5}.
\newblock \urlprefix\url{https://doi.org/10.1038/s41550-018-0562-5}

\bibitem{2006SoPh..234...41K}
J.A. {Klimchuk}, \solphys \textbf{234}(1), 41 (2006).
\newblock \doi{10.1007/s11207-006-0055-z}

\bibitem{williams2004soho}
D.R. Williams, in \emph{{SOHO 13}---Waves, Oscillations and Small-Scale Transient Events in the Solar Atmosphere: A Joint View from SOHO and TRACE}, ed. by H.~Lacoste, {ESA SP-547} (European Space Agency (ESA) Publications Division, Noordwijk, The Netherlands, 2004), pp. 513--518

\bibitem{1999SoPh..188...89C}
R.~{Cowsik}, J.~{Singh}, A.K. {Saxena}, R.~{Srinivasan}, A.V. {Raveendran}, \solphys \textbf{188}(1), 89 (1999).
\newblock \doi{10.1023/A:1005149303094}

\bibitem{1996PASJ...48..545I}
K.~{Ichimoto}, K.~{Kumagai}, I.~{Sano}, T.~{Kobiki}, A.~{Munoz}, T.~{Sakurai}, \pasj \textbf{48}(3), 545 (1996).
\newblock \doi{10.1093/pasj/48.3.545}

\bibitem{2003ApJ...599..596R}
N.L. {Reginald}, O.C.S. {Cyr}, J.M. {Davila}, J.W. {Brosius}, \apj \textbf{599}(1), 596 (2003).
\newblock \doi{10.1086/379148}

\bibitem{Hanaoka:2021aa}
Y.~Hanaoka, Y.~Sakai, K.~Takahashi, Solar Physics \textbf{296}(11), 158 (2021).
\newblock \doi{10.1007/s11207-021-01907-0}.
\newblock \urlprefix\url{https://doi.org/10.1007/s11207-021-01907-0}

\bibitem{Hanaoka2024}
Y.~Hanaoka, Y.~Sakai, Y.~Masuda, Frontiers in Astronomy and Space Sciences \textbf{11} (2024).
\newblock \doi{10.3389/fspas.2024.1458746}

\bibitem{Druckmuller2024}
M.~Druckm{\"u}ller, V.~Ru{\v{s}}in, P.~Aniol, S.R. Habbal, The Astrophysical Journal Letters \textbf{965}(1), L10 (2024).
\newblock \doi{10.3847/2041-8213/ad353b}

\bibitem{2020ApJ...904..178B}
A.~{Bemporad}, \apj \textbf{904}(2), 178 (2020).
\newblock \doi{10.3847/1538-4357/abc482}

\bibitem{2023ApJ...946...14B}
A.~{Bemporad}, \apj \textbf{946}(1), 14 (2023).
\newblock \doi{10.3847/1538-4357/acb8b8}

\bibitem{Boe2020}
B.~Boe, S.~Habbal, M.~Druckm{\"u}ller, The Astrophysical Journal \textbf{895}(2), L35 (2020).
\newblock \doi{10.3847/1538-4357/ab8ae6}.
\newblock \urlprefix\url{https://doi.org/10.3847/1538-4357/ab8ae6}

\bibitem{10.1063/1.1618540}
R.~Woo, S.R. Habbal, AIP Conference Proceedings \textbf{679}(1), 55 (2003).
\newblock \doi{10.1063/1.1618540}.
\newblock \urlprefix\url{https://doi.org/10.1063/1.1618540}

\bibitem{Zeiler_World_Eclipses_2040}
M.~Zeiler.
\newblock World eclipses: 2021 to 2040 (2025).
\newblock \urlprefix\url{https://www.eclipse-maps.com/}.
\newblock (Map of world eclipses from 2021 to 2040)

\bibitem{M.-Pasachoff:2017aa}
J.~M.~Pasachoff, Nature Astronomy \textbf{1}(8), 0190 (2017).
\newblock \doi{10.1038/s41550-017-0190}.
\newblock \urlprefix\url{https://doi.org/10.1038/s41550-017-0190}

\bibitem{Liberatore:2022aa}
A.~Liberatore, G.~Capobianco, S.~Fineschi, G.~Massone, L.~Zangrilli, R.~Susino, G.~Nicolini, Solar Physics \textbf{297}(3), 29 (2022).
\newblock \doi{10.1007/s11207-022-01958-x}.
\newblock \urlprefix\url{https://doi.org/10.1007/s11207-022-01958-x}

\bibitem{ShelleyCameraSummary}
M.~Shelley.
\newblock Dslr/mirrorless camera artefact summary.
\newblock \url{https://markshelley.co.uk/Astronomy/camera_summary.html}.
\newblock Accessed: 2025-10-20

\bibitem{AstronomyToolsFOV}
{Astronomy.tools}.
\newblock Field of view calculator.
\newblock \url{https://astronomy.tools/calculators/field_of_view/}.
\newblock Accessed: 2025-10-20

\bibitem{Molnar:2025aa}
M.E. Molnar, R.~Casini, P.~Bryans, B.~Berkey, K.~Tyson, Solar Physics \textbf{300}(7), 88 (2025).
\newblock \doi{10.1007/s11207-025-02500-5}.
\newblock \urlprefix\url{https://doi.org/10.1007/s11207-025-02500-5}

\bibitem{HutechHinode}
{Hutech Astronomical Products}.
\newblock Hinode solar guider.
\newblock \url{https://www.sciencecenter.net/hutech/Hinode-sg/index.htm}.
\newblock Accessed: 2025-10-20

\bibitem{Phil2024}
P.~Hart (2024).
\newblock Private communication

\bibitem{2007tseh.book.....M}
M.~{Mobberley}, \emph{{Total Solar Eclipses and How to Observe Them}} (2007).
\newblock \doi{10.1007/978-0-387-69828-1}

\bibitem{1836MNRAS...4...15B}
F.~{Baily}, \mnras \textbf{4}, 15 (1836).
\newblock \doi{10.1093/mnras/4.2.15}

\bibitem{JubierCircumstancesCalc}
X.M. Jubier.
\newblock Local circumstances calculator (v1.0.6).
\newblock \url{http://xjubier.free.fr/en/site_pages/SolarEclipseCalc_Diagram.html} (2007).
\newblock Last page update on July 7, 2007. Accessed: 2025-10-20

\bibitem{JubierExposureCalc}
X.M. Jubier.
\newblock Shutter speed calculator for solar eclipses (v1.0.2).
\newblock \url{http://xjubier.free.fr/en/site_pages/SolarEclipseExposure.html} (2017).
\newblock Last page update on July 3, 2017. Accessed: 2025-10-20

\bibitem{JubierMaestro}
X.M. Jubier.
\newblock Solar eclipse maestro for macos x.
\newblock \url{http://xjubier.free.fr/en/site_pages/solar_eclipses/Solar_Eclipse_Maestro_Photography_Software.html} (2021).
\newblock Last page update on November 9, 2021. Accessed: 2025-10-20

\bibitem{Edwards:2023aa}
L.~Edwards, K.A. Bunting, B.~Ramsey, M.~Gunn, T.~Fearn, T.~Knight, G.D. Muro, H.~Morgan, Solar Physics \textbf{298}(12), 140 (2023).
\newblock \doi{10.1007/s11207-023-02231-5}.
\newblock \urlprefix\url{https://doi.org/10.1007/s11207-023-02231-5}

\bibitem{Voulgaris:2012aa}
A.G. Voulgaris, P.S. Gaintatzis, J.H. Seiradakis, J.M. Pasachoff, T.E. Economou, Solar Physics \textbf{278}(1), 187 (2012).
\newblock \doi{10.1007/s11207-012-9929-4}.
\newblock \urlprefix\url{https://doi.org/10.1007/s11207-012-9929-4}

\bibitem{Voulgaris:2022aa}
A.G. Voulgaris, C.~Mouratidis, K.~Tziotziou, J.H. Seiradakis, J.M. Pasachoff, Solar Physics \textbf{297}(4), 49 (2022).
\newblock \doi{10.1007/s11207-022-01983-w}.
\newblock \urlprefix\url{https://doi.org/10.1007/s11207-022-01983-w}

\bibitem{2008ApOpt..47.5744N}
M.W. {Noble}, D.M. {Rust}, P.N. {Bernasconi}, J.M. {Pasachoff}, B.A. {Babcock}, M.A. {Bruck}, \ao \textbf{47}(31), 5744 (2008).
\newblock \doi{10.1364/AO.47.005744}

\bibitem{Muro:2023aa}
G.D. Muro, M.~Gunn, S.~Fearn, T.~Fearn, H.~Morgan, Solar Physics \textbf{298}(6), 75 (2023).
\newblock \doi{10.1007/s11207-023-02162-1}.
\newblock \urlprefix\url{https://doi.org/10.1007/s11207-023-02162-1}

\end{thebibliography}

\end{document}